\documentclass[journal=jacsat,manuscript=article]{achemso}

\usepackage[version=3]{mhchem} 
\usepackage{booktabs}
\usepackage{multirow}
\usepackage{hyperref}
\usepackage{soul}
\usepackage{xcolor}
\usepackage{bm}
\usepackage{natbib}

\author{Priyanka Pandey}
\affiliation{Department of Chemistry and Cherry L. Emerson Center for Scientific Computation, Emory University, Atlanta, Georgia 30322, U.S.A.}

\author{Mrinal Arandhara}
\affiliation{Department of Inorganic and Physical Chemistry, Indian Institute of Science, Bangalore 560012, India}

\author{Paul L. Houston}
\affiliation{Department of Chemistry and Chemical Biology, Cornell University, Ithaca, New York
14853, U.S.A. and Department of Chemistry and Biochemistry, Georgia Institute of
Technology, Atlanta, Georgia 30332, U.S.A}
\email{plh2@cornell.edu}

\author{Chen Qu}
\affiliation{Independent Researcher, Toronto, Ontario M9B0E3, Canada}

\author{Riccardo Conte}
\affiliation{Dipartimento di Chimica, Universit\`a degli Studi di Milano, 20133 Milano, Italy}

\author{Joel M. Bowman}
\email{jmbowma@emory.edu}
\affiliation{Department of Chemistry and Cherry L. Emerson Center for Scientific Computation, Emory University, Atlanta, Georgia 30322, U.S.A.}

\author{Sai G. Ramesh}
\email{sairamesh@iisc.ac.in}
\affiliation{Department of Inorganic and Physical Chemistry, Indian Institute of Science, Bangalore 560012, India}

\title{Assessing PIP and sGDML Potential Energy Surfaces for \ce{H3O2-}} 
\abbreviations{PES ML}

\begin{document}

\begin{tocentry}
\begin{center}
\includegraphics[width=\textwidth]{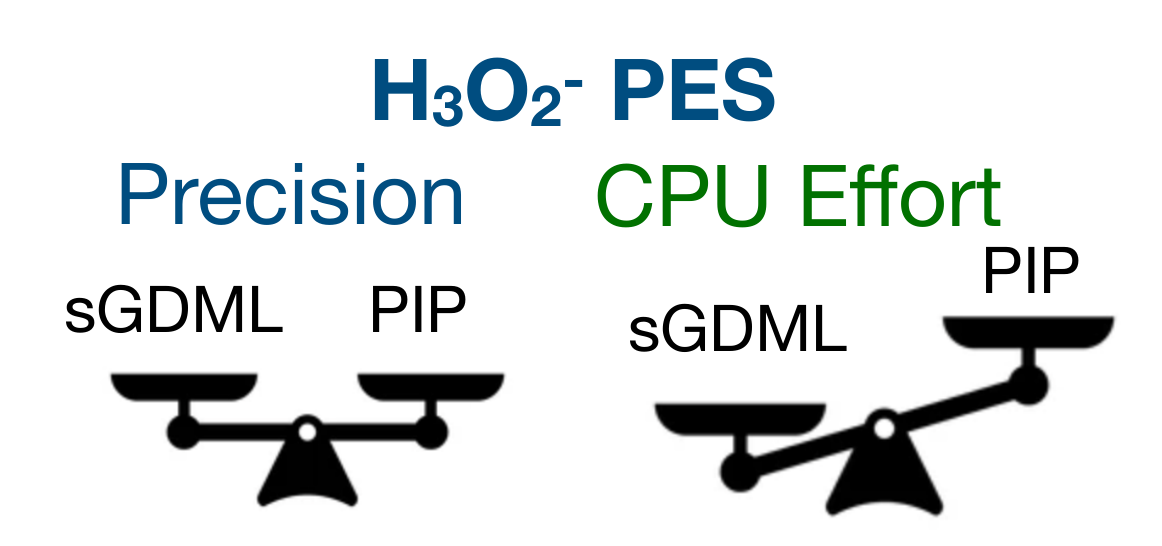}
\end{center}
\end{tocentry}

\newpage
\begin{abstract}
The singly hydrated hydroxide anion \ce{OH^-(H2O)} is of central importance to a detailed molecular understanding of water, so there is strong motivation to develop a highly accurate potential to describe this anion.  While this is a small molecule, it is necessary to have an extensive dataset of energies and, if possible, forces to span several important stationary points. Here we assess two machine-learned potentials,  one using the symmetric gradient domain machine learning (sGDML) method and one based on permutationally invariant polynomials (PIPs).  These are successors to a PIP potential energy surface (PES) reported in 2004.  We describe the details of both fitting methods and then compare the two PESs with respect to precision, properties, and speed of evaluation.   While the precision of the potentials is similar, the PIP PES is much faster to evaluate for energies and energies plus gradient than the sGDML one. Diffusion Monte Carlo calculations of the ground vibrational state, using both potentials, produce similar large anharmonic downshift of the zero-point energy compared to the harmonic approximation  the PIP and sGDML potentials.  The computational time for these calculations using the sGDML PES is roughly 300 times greater than using the PIP one. 
\end{abstract}

\section{Introduction}
The singly hydrated hydroxide anion \ce{OH^-(H2O)} has long been of interest to theorists and experimentalists.\cite{tuckerman2002nature,neumark2002spectroscopy,huang2004quantum,lee2004structures,mccoy2005quantum,diken2005fundamental,Tachikawa2005,yang2008full,Suzuki2008,McCoy2009,pelaez2017infrared,pmid:12548680}  The first \textit{ab initio}-based, full-dimensional, machine-learned potential energy (MLP) was reported in 2004\cite{huang2004quantum,mccoy2005quantum} using permutationally invariant polynomials (PIPs) in terms of primary and secondary PIPs.\cite{Braams2009} In brief, this PES was a least-squares fit to almost 67 000 \textit{ab initio} energies \cite{huang2004quantum} (later updated to a fit to about 23 000 energies\cite{mccoy2005quantum}), obtained with the CCSD(T) method with an aug-cc-pVTZ basis. The variables of the fit are the ten internuclear distances, and the polynomial basis is constructed to be permutationally invariant with respect to the permutation of like atoms.  This PIP PES was used in VSCF/VCI (reaction path) and fixed-node diffusion Monte Carlo calculations of vibrational energies. While this PES was successful in obtaining these energies and making insightful comparisons with experiment, it did not have extensive coverage of the high-energy saddle point for the exchange of the shared H-atom with the terminal one, also referred to as the bifurcation saddle point. The structure of this saddle point as well as the global minimum and H-atom transfer saddle point are shown in Fig. \ref{fig:saddle_config} below. 
\begin{figure}
    \centering
    \includegraphics[width=0.7\textwidth]{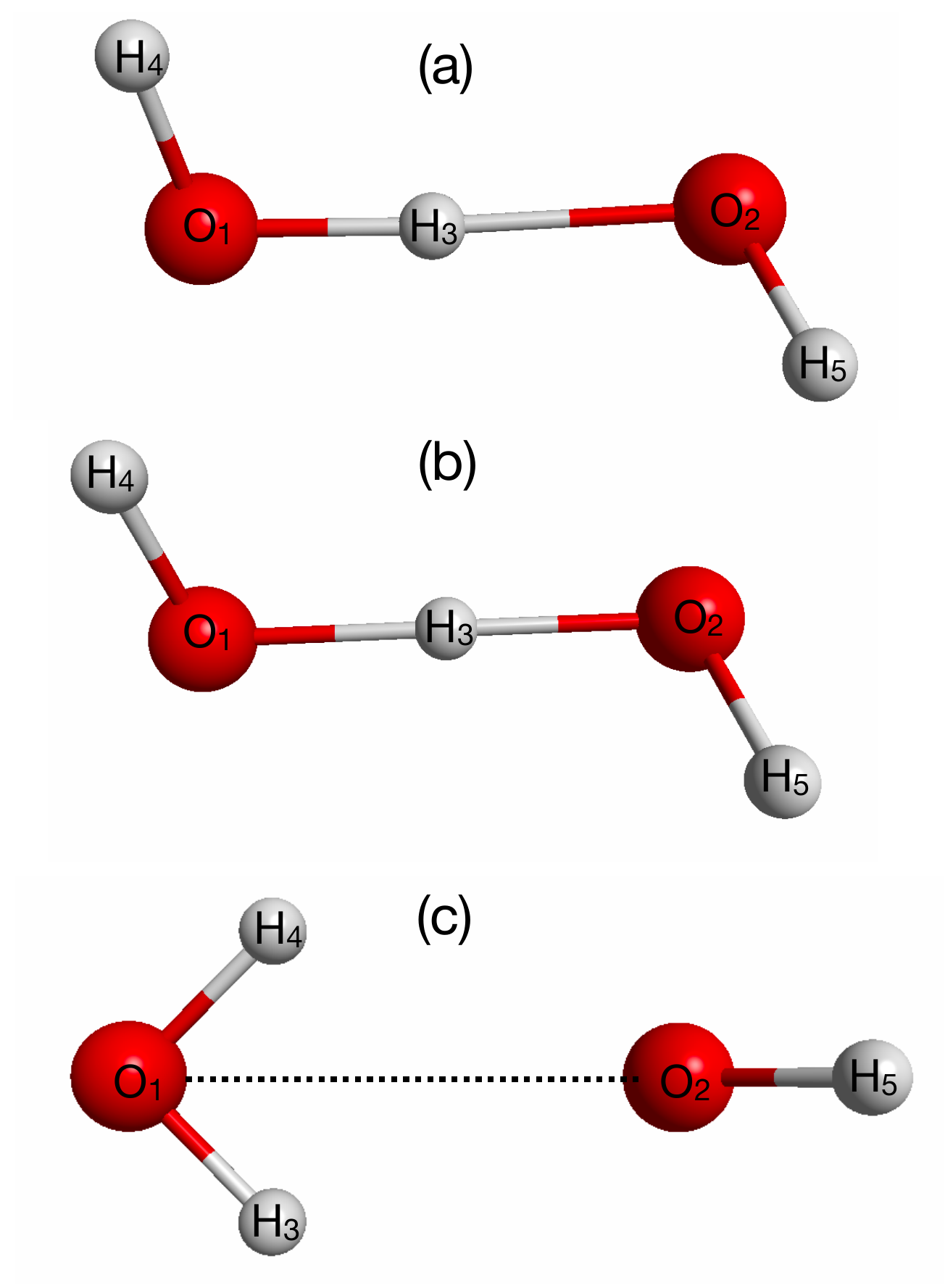}
    \caption{The structures of three stationary points of \ce{H3O2-} (a) global minimum, (b) H-transfer TS, (c) bifurcation TS }
    \label{fig:saddle_config}
\end{figure}

This lack of coverage was remedied very recently by two of us (MA and SGR)\cite{arandhara24} who calculated 15 024 energies and gradients at the CCSD(T)/aVTZ level of theory using CFOUR.\cite{cfour}  They trained a symmetric gradient domain machine learning (sGDML)\cite{Chmiela2017, Chmiela2018, sgdml2019, Sauceda2019, Sauceda2020} PES on a subset of 3000 energies and gradients, with validation (hyperparameter adjustment) using another 3000 data points, and then tested it on the remaining data. The PES was then used in path integral simulations of the temperature dependence of the motion along the so-called bifurcation pathway, whose TS is shown in Fig.~\ref{fig:saddle_config}.  

 This dataset and the sGDML PES provide an opportunity to assess that PES and a PIP PES trained on this dataset.  We do that here.  In the next section, we provide details of the sGDML and PIP approaches and the specifics for this particular dataset.  The performance of the two fits is examined in detail in the Results and Discussion section. Results of diffusion Monte Carlo calculations are also presented in that section. The final section contains a summary and conclusions.

\newpage
\section{Fitting Methods}
\subsection{Permutationally Invariant Polynomials and MSA Software}
MLPs using a basis of permutationally invariant polynomials (PIPs) have been reported for nearly 20 years, with the PES for \ce{H3O2-} in 2004 being one of the first such MLPs. The expression for the PIP potential is given by

\begin{equation}
V(\mathbf{y})= \sum_{i=1}^{n_p} c_i p_i(\mathbf{y}),
\label{eq1}
\end{equation}

\noindent
where $c_i$ are linear coefficients, $p_i$ are PIPs, $n_p$ is the total number of polynomials (and linear coefficients $c_i$) for a given maximum polynomial order, and $\mathbf{y}$ are transformed internuclear distances.  We have used the following 3 transformations: $y_{ij}=\exp(-r_{ij}/a)$, $y_{ij}=\exp(-r_{ij}/a)/r_{ij}$,\cite{Braams2009} and $y_{ij}$=$1/r_{ij}$\cite{wang2016five}. PIPs are polynomials that are invariant with respect to permutations of like atoms. 

Our current software to generate PIPs and perform least squares fitting\cite{msachen, PESPIP} is based on monomial symmetrization \cite{Braams2009, Xie10}.   The first part of MSA creates the PIP basis and writes it to a text file named ``basis.f90", where the file has the suffix .f90 for later use in Fortran. Analytical gradients are also provided with this code.  In a second step, fast gradient evaluation is available, based on reverse differentiation algorithms, and Mathematica scripts. These are described in detail elsewhere\cite{PESPIP} and are used here. The final code is written in Fortran 90.

To complete this short review, we note that the linear coefficients $c_i$ are optimized to minimize the L2 loss, i.e., the sum of the square of the differences between the data and fit $V(\mathbf{y};\mathbf{c})$, where we explicitly indicate the parametric dependence on the coefficients $\mathbf{c}$.  The standard approach leads to the matrix equation
\begin{equation}
    \mathbf{Ac}= \mathbf{d},
    \label{eq2}
\end{equation}
where the matrix $\mathbf{A}$, elements of which are given by $A_{i,j}$ = $p_j(\mathbf{y}_i)$, is $N \times n_p$,  where $N$ is the size of the dataset of energies plus gradients (if they are used), $\mathbf{c}$ is the column vector of length $n_p$ and $\mathbf{d}$ is the column vector length $N$ and consists of these data.  In general, $n_p \ll N$, and so this is an overdetermined set of linear equations. 
The solution to this least-squares problem is given formally by
\begin{equation}
     \mathbf{c}=(\mathbf{A}^T\mathbf{A})^{-1}(\mathbf{A}^T\mathbf{d}).
     \label{eq3}
\end{equation}
There are several ways to proceed; we use singular value decomposition of the matrix $\mathbf{A} = \mathbf{U} \boldsymbol{\Sigma} \mathbf{V}^T$, where $\mathbf{U}$ and $\mathbf{V}$ are orthogonal matrices of size $N \times N$ and $n_p \times n_p$, respectively, and  
$\boldsymbol{\Sigma}$ is a diagonal matrix of $n_p$ singular values in descending order with zeros below the diagonal element $\sigma_{n_p}$.
$\mathbf{U}$ can be partitioned into two blocks, $\mathbf{U}_1$ and $\mathbf{U}_2$, where $\mathbf{U}_1$ is $N \times n_p$. The final expression for the coefficients is
\begin{equation}
    \mathbf{c}=(\mathbf{V} \boldsymbol\Sigma^{-1} \mathbf{U}_1^T)\mathbf{d}.
    \label{eq4}
\end{equation}
We use dgelss.f90 for this analysis.

In general, we fit an entire data set (including gradients if available).  This is because the linear regression method depends on the size of the PIP basis and not the size of the dataset. So, bases with $n_p$ of the order of thousands present no difficulties for data sets that are an order of magnitude bigger.  Of course testing of the fit is done, generally on out-of-sample data. This protocol is not the usual split-train-test protocol.  Methods based on Kernel Ridge Regression and Gaussian Process Regression are ``trained" directly on a dataset and the resulting linear algebra problem, i.e., a matrix inverse of the Kernel at the configurations of the dataset, is limited to datasets of the order of thousands, much smaller than the datasets used in PIP Linear Regression.
\begin{figure}[htbp!]
\includegraphics[width=0.8\columnwidth]{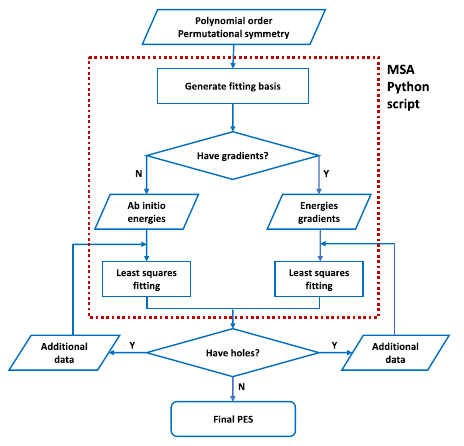}
\caption{Flow chart of the PES fitting procedures. The procedures in the red rectangle are now integrated in one single Python script.}
\label{fig:flow}
\end{figure}

The overall work flow of the MSA software to obtain a PIP PES is given in Figure \ref{fig:flow}.  In the present work, energies and gradients are included in the dataset. DMC calculations are run to locate large negative regions of the fit, ``holes".  In general, the holes occur at \st{actual} high energies, which (not surprisingly) were not sampled in the dataset. Additional data are added at the hole configuration and a new fit is done. This is repeated until there are no or very few holes.

We provide specific details of this approach below where we discuss the fit to the \ce{H3O2-} dataset.

\subsection{sGDML}

Symmetric Gradient domain machine learning (sGDML) refers to a kernel ridge method that fits the gradient of the potential while identifying and incorporating symmetries of the molecule.\cite{Chmiela2017, Chmiela2018, sgdml2019, Sauceda2019, Sauceda2020} This approach has been widely used for numerous applications and the Python code for usage is available.\cite{sgdmlcode} Two of us (MA and SR), who developed the sGDML PES for \ce{H3O2-} trained the model using the Python code but wrote a Fortran 90 code to evaluate the gradient and energy. The sGDML potential and Fortan code are available on \href{https://github.com/arandharamrinal}{Github}.

We briefly describe how sGDML works. The input to sGDML is a set of molecules geometries, their ab initio energies and atomic forces obtained from a high-level quantum chemistry calculation. As per the choice of the user, this data set is split into training, validation and test points. The points in each set are so chosen that the energy distribution in each of the three data sets is consistent with the distribution of the full data set. From (a subset of) the set of $M$ training points, sGDML identifies the molecular symmetries in a data-driven manner \cite{Chmiela2018, sgdml2019}, producing atomic permutations matrices. In the case of {H$_3$O$_2^-$}, all 12=2!3! permutations are found. A set of descriptors comprising of all $N(N-1)/2$ inverse pairwise interatomic distances are constructed for each training set geometry (10 in the present case). Using this, a kernel matrix is constructed in descriptor space where all identified permutational symmetries summed over. This is finally transformed back to Cartesian space yielding a $3NM \times 3NM$ matrix, $\mathbf{K}$. Using the atomic forces as a vector $\mathbf{f}$ of length $3NM$, the equation
\begin{equation}
(\mathbf{K} + \lambda \mathbf{I}) \mbox{\boldmath$\alpha$} = \mathbf{-f}
\label{eq:sgdml_fit}
\end{equation}
is solved, where $\lambda$ is a regularization parameter and $\mathbf{I}$ is the identity matrix. 

The coefficient vector $\mathbf{\alpha}$ is the main result of the training. The values of the coefficients depend on a hyperparameter $\sigma$ that in used in the Mat{\'e}rn kernel. In order to choose the optimal $\sigma$, the error is the evaluated using the validation data set (rather than the training set) and $\sigma$ is changed on a grid (e.g. in units of 1), and Eq.~\eqref{eq:sgdml_fit} is solved again with each new $\sigma$ until the error is the least. The optimal $\sigma$ and $\mathbf{\alpha}$ are then using for testing and for prediction.

At a new geometry $\mathbf{x}$, the $3N$ forces $\hat{\mathbf{f}}(\mathbf{x})$ at a new configuration is carried out by evaluating
\begin{equation}
\hat{\mathbf{f}}(\mathbf{x}) = \sum_i \mathbf{K}(\mathbf{x},\mathbf{x}_i) \alpha_i
\label{eq:sgdml_pred}
\end{equation}
where the summation runs over all the training points and their replicas through permutations. The fitting coefficients are also suitably permuted; they carry the same symmetry as the data points. A technical aspect is that for the prediction stage, the $\mathbf{\alpha}$ are saved in descriptor space and the the kernel matrix between the query and training points is also prepared in this space. Hence, the forces obtained are first obtained in the descriptors (inverse distances) and then transformed by the chain rule to Cartesian space. Eq.~\eqref{eq:sgdml_pred} is used to both obtain force and energy errors in the test data set as well as the predict them at a queried geometry. The energies are obtained as an integral over the forces.

We note in passing that a possible alternative to the inverse distance descriptors is the use PIPs in sGDML, which build in the symmetry aspect of the ML potential.  This approach has been used with  great success for Neural Network potentials~\cite{PIP-NN-1,NNZhang16} and also for Gaussian Approximation potentials.\cite{PIP-GP} 

\section{Results and discussion}
\subsection{\ce{H3O2-} dataset}

\begin{figure}[htbp!]
    \centering
    \includegraphics[width=1.0\textwidth]{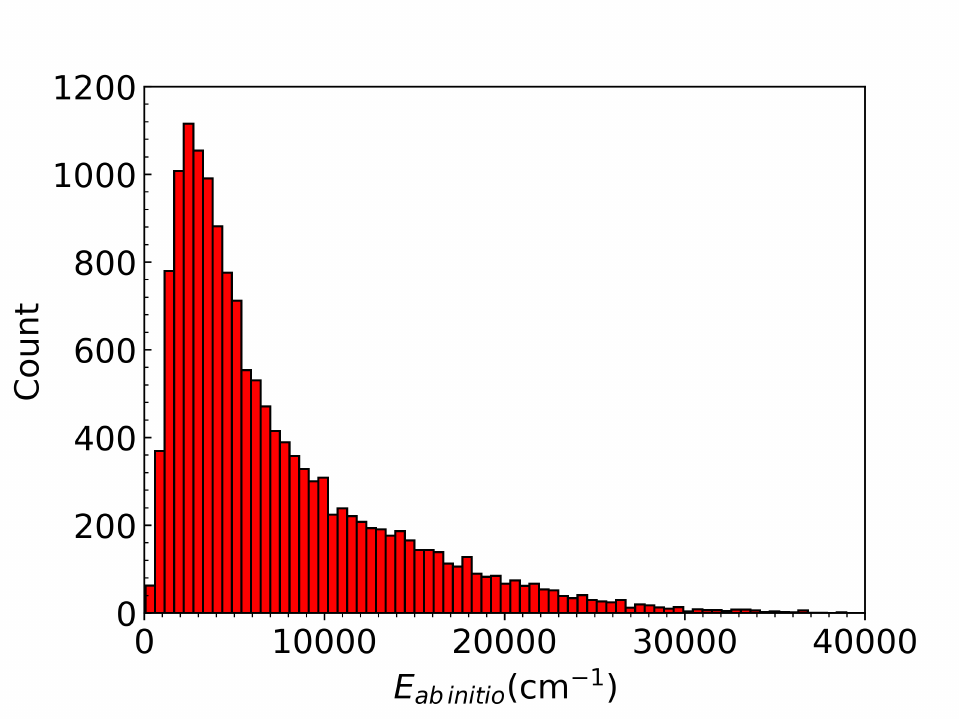}
    \caption{Histogram of the \textit{ab initio} energies in the full dataset. See Ref.~\cite{arandhara24} for the construction of the dataset.}
    \label{fig:histofenergies}
\end{figure}

A detailed description of the dataset of 15 024  CCSD(T) energies and gradients has been reported \cite{arandhara24}, so we just briefly summarize it here.  The energies extend to 54 000 cm$^{-1}$ with a concentration of energies at roughly 10 000 cm$^{-1}$. The distribution of the energies is shown in the Figure~\ref{fig:histofenergies}. As seen, most of the energies are below 20 000 cm$^{-1}$ with a small number extending to 50 000 cm$^{-1}$, beyond the range of abscissa. 

The complete dataset of energies and gradients is not practical for training and prediction with the sGDML method.  Thus,  3000  configurations and 45 000 gradients were selected for training, with another 3000 points used for validation through which the hyperparameter is adjusted. Testing was performed using the remaining 9024 points. Further details of the sGDML fit have been given previously\cite{arandhara24} and so we do not repeat those here. For the PIP fit the total dataset was used with no weighting of the data. Another fit was done using 6000 configurations, so a total data size of 96 000. These are the same points used in the sGDML training and validation steps.

\subsection{Precision and performance of the fits}
The new PIP PESs use full permutational symmetry, i.e., 3!2! = 12 and a maximum polynomial order of seven.  This results in a basis size of 2022 PIPs.  The generation of this basis and the fitting are both fast (about 5 mins of wall-clock time) using MSA software~\cite{msachen}. The energies and gradients are not weighted, as noted already, and the Morse range parameter equals 3 bohr. A short video showing the interactive steps to do this on a linux workstation in command line mode can be found 
\href{https://scholarblogs.emory.edu/bowman/msa/}{\textbf{here}}. 

\begin{figure}[htbp!]
    \centering
    \includegraphics[width=1.0\columnwidth]{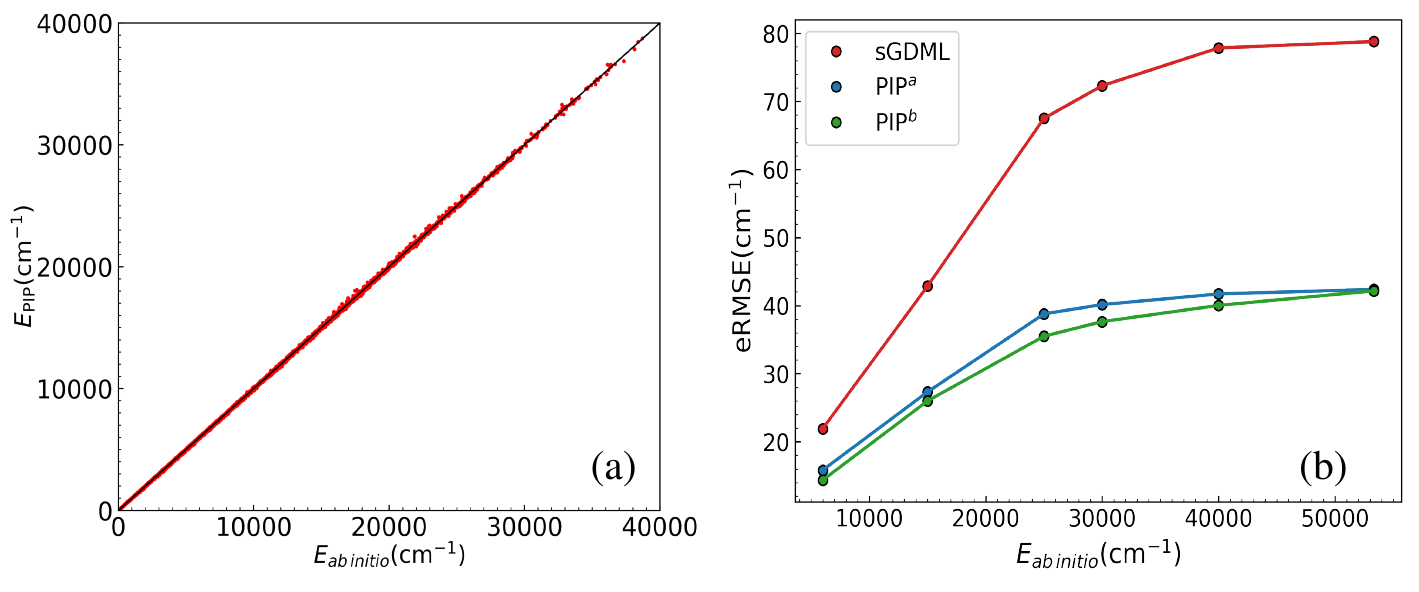}
    \caption{(a) Correlation between the PIP PES fit to all data and CCSD(T)/aVTZ energies. (b) eRMSE for PIP$^a$, fit to the full data set, PIP$^b$, fit to data at 6000 configurations,  and sGDML PES  as a function of the \textit{ab initio} energy.
    }
    
    \label{fig:correlation}
\end{figure}

To begin the assessment of the new PIP PES, we show in Figure~\ref{fig:correlation} the correlation plot of the PIP fit and the eRMSE vs energy for all 15 024 energies. Also shown is the eRMSE for the PIP PES trained on 6000 configurations, denoted PIP$^b$, and the sGDML PES.  As seen, the PIP PES eRMSEs are about half those of sGDML.  Very high precision is seen for energies up to 20 000 cm$^{-1}$, which is sufficient even for quantum studies of the dynamics of this complex. 

The energy and force RMSEs for the PIP and sGDML PESs are given in Table \ref{tab:rmse}. The RMSEs for the PIP fits using the full and a subset of the data are almost the same and are smaller than the corresponding ones already reported for sGDML\cite{arandhara24}. It is perhaps noteworthy that the PIP PES fRMSE is lower than the sGDML one.  

\begin{table}[!tb]
\renewcommand{\arraystretch}{1.5}
   \centering
   \begin{tabular}{l l r r r} \hline
        Error & Units  & sGDML & PIP$^a$ & PIP$^b$  \\ \hline \hline
         eRMSE & cm$^{-1}$& $78.8$ & $42.4$ & $42.2$\\
         fRMSE & cm$^{-1}$ \r{A}$^{-1}$ dof$^{-1}$ & $194.1$ & $141.4$ & $131.5$\\
         \hline
         $^a$Fit using all data\\
         $^b$Fit using data at 6000 configurations\\
             \caption{The RMS errors in fitted energies and forces computed using all the CCSD(T)/aug-cc-pVTZ energies (E) and forces (F).}
    \label{tab:rmse}
    \end{tabular}
\end{table}

Next, we consider some properties of the various PESs.  Hereafter, we consider only the PIP PES fit to all the data. First, we show the internal coordinates of the global minimum in Table \ref{tab:Opt_geom}.  As seen, both PESs are in good agreement with the direct \textit{ab initio} values.  This is gratifying since the \textit{ab initio} minimum configuration is not included in the training dataset.
\begin{table}[!tb]
\renewcommand{\arraystretch}{1.5}
    \centering
    \begin{tabular}{l r r r} \hline
                       & \textit{Ab initio} &   sGDML  & PIP  \\ \hline \hline
         R(${\ce{O1O2}}$) &   $2.4887$  & $2.4836$ & $2.4840$ \\
         R(\ce{O1H3})  &   $1.0904$  & $1.0844$ & $1.0842$\\
         R(\ce{O1H4})  &   $0.9614$  & $0.9588$ & $0.9588$\\
         R(\ce{O2H5})  &   $0.9641$  & $0.9612$ & $0.9611$\\
         $\theta$(\ce{H3O1O2}) & $1.48$   &  $1.50$  & $1.47$\\ 
         $\theta$(\ce{H4O1O2}) & $100.44$ & $100.46$ &  $100.41$ \\ 
         $\theta$(\ce{H5O2O1}) & $105.72$ & $105.83$ & $105.81$ \\ 
         $\phi$(\ce{H3O1O2H5}) & $-61.66$ & $-62.18$  & $-62.83$\\ 
         $\phi$(\ce{H4O1O2H5}) & $102.26$ &  $101.37$ & $100.85$\\ 
         \hline
    \end{tabular}
    \caption{Optimized structure of \ce{H3O2-} global minimum. All bond lengths are in \r{A} while the bond angles $\theta$ and dihedral angles $\phi$ are in degrees.}
    \label{tab:Opt_geom}
\end{table}
Next consider harmonic frequencies at the global minimum; these are give in Table \ref{tab:freq_mode}.  Here again, the two PESs perform almost equally, with the exception of lowest frequency mode, where the PIP PES is more precise.

\begin{table}[!tb]
\renewcommand{\arraystretch}{1.5}
    \centering
    \begin{tabular}{l r r r} \hline
             Mode & \textit{Ab initio} &sGDML  & PIP   \\ \hline \hline
        \ce{HOOH} dih     &  203.0  & 158.4  & 193.0 \\
        \ce{O-O} str      &  326.1  & 316.5  & 322.3 \\
        \ce{OH-} bend     &  470.4  & 467.6  & 482.3 \\
        \ce{H2O} rock     &  585.9  & 583.6  & 590.9 \\
    \ce{OH_{in}} oop bend &  1354.9 & 1353.9 & 1366.3 \\
        \ce{OH_{in}} str  &  1605.1 & 1627.4 & 1598.8 \\
    \ce{OH_{in}} i.p. bend&  1739.2 & 1756.7 & 1734.9 \\
        \ce{OH-} str      &  3815.0 & 3804.4 &  3814.6 \\
        \ce{OH_{out}} str &  3866.3 & 3863.1 &  3866.3\\ \hline

    \end{tabular}
    \caption{Comparison of harmonic frequencies (in cm$^{-1}$) at the optimized global minimum geometry from \textit{ab initio} calculations at CCSD(T)/aug-cc-pVTZ level of theory using CFOUR $2.1$\cite{cfour}, the sGDML and PIP PESs.}
    \label{tab:freq_mode}
\end{table}

Next, we consider the various saddle points. A comparison of \textit{ab initio} normal mode frequencies of \ce{H3O2-} with those from both PESs at the various stationary points, along with their energies, is presented in the Table~\ref{tab:freq_mode_ts1} and \ref{tab:freq_mode_ts2}. Note that the energies are relative to the global minimum energy. The Cartesian coordinates of these saddle points along with the global minimum, obtained from the PIP PES optimizations, are given in the supplementary information.

\begin{table}[!tb]
\renewcommand{\arraystretch}{1.5}
    \centering
    \begin{tabular}{l r r r r r r r r r r r r} 
     \toprule
             \multirow{2}{*}{Mode} & \multicolumn{3}{c}{TS Bifurcation} & \multicolumn{3}{c}{TS H-transfer}\\
              \cmidrule(lr){2-4}  \cmidrule(lr){5-7}
             &  \textit{Ab initio} &  sGDML  & PIP  &  \textit{Ab initio} &sGDML  & PIP \\ 
    \midrule
              Q$_1$ &  443.7i & 509.5i & 471.9i & 667.8i & 705.1i & 642.6i \\ 
              Q$_2$ &  259.6 &  156.9  & 171.5 & 210.6 & 171.8 & 202.3\\ 
              Q$_3$ &  287.9 &  275.4  &289.1 & 568.5 & 566.0 & 576.43\\ 
              Q$_4$ &  390.7 &  295.8  &346.5 & 577.4 & 575.7 & 579.41\\ 
              Q$_5$ &  880.2 &  853.4  &850.9 & 632.2 & 630.4 & 636.6\\ 
              Q$_6$ &  1658.8 & 1651.6 &1659.1 & 1528.1 & 1524.2& 1532.7\\ 
              Q$_7$ &  3613.1 & 3653.2 &3685.1 & 1626.6 & 1643.4& 1630.4\\ 
              Q$_8$ &  3676.4 & 3749.8 &3736.4 & 3840.0 & 3842.4& 3840.6\\ 
              Q$_9$ &  3757.9 & 3810.1 &3799.8 & 3840.6 & 3847.3 & 3840.9\\ 
              \hline
              E &  2521.1 & 2552.9 & 2478.4 & 81.0  & 85.3 & 74.8\\ 
              \hline
    \end{tabular}
    \caption{Comparison of normal mode frequencies and energies (in cm$^{-1}$) of \ce{H3O2-} at the bifurcation TS and the shared proton transfer TS, from \textit{ab initio} calculations at the CCSD(T)/aug-cc-pVTZ level using CFOUR 2.1, sGDML PES and PIP PES.} 
    \label{tab:freq_mode_ts1}
\end{table}

\begin{table}[]
\renewcommand{\arraystretch}{1.5}
    \centering
    \begin{tabular}{l r r r r r r r r r r r r} 
     \toprule
             \multirow{2}{*}{Mode} & \multicolumn{3}{c}{TS $cis$} & \multicolumn{3}{c}{TS $trans$}\\
              \cmidrule(lr){2-4}  \cmidrule(lr){5-7}
             & \textit{Ab initio} &  sGDML  & PIP  &  \textit{Ab initio} &sGDML  & PIP \\ 
    \midrule
              Q$_1$ & 229.2i &  205.9i & 224.3i & 182.4i & 122.2i & 161.2i\\ 
              Q$_2$ &  322.5 &  326.0  & 321.4  & 312.1  & 311.2 & 311.9\\ 
              Q$_3$ &  437.8 &  461.2  & 454.3  & 413.7  & 417.5 & 425.2\\ 
              Q$_4$ &  676.2 &  691.0  & 690.4  & 690.7  & 695.3 & 700.8\\ 
              Q$_5$ & 1178.7 &  1179.7 & 1182.1 & 1181.5 & 1180.0 & 1193.6\\ 
              Q$_6$ & 1713.1 &  1729.0 & 1725.5 & 1696.9 & 1726.5 & 1688.4\\ 
              Q$_7$ & 1838.1 &  1838.2 & 1842.2 & 1817.1 & 1834.5 & 1808.3\\ 
              Q$_8$ & 3809.5 &  3805.9 & 3810.8 & 3819.0 & 3799.8 & 3820.0\\ 
              Q$_9$ & 3868.2 &  3860.9 & 3874.0 & 3867.8 & 3867.3 & 3869.9\\ 
              \hline
              E &  373.6& 309.0 & 357.1 & 165.4 & 75.2 & 130.2\\ 
              \hline
    \end{tabular}
    \caption{Comparison of normal mode frequencies and energies (in cm$^{-1}$) of \ce{H3O2-} at the $cis$ and $trans$ \ce{HO-OH} torsion barriers, from \textit{ab initio} calculations at the CCSD(T)/aug-cc-pVTZ level using CFOUR 2.1, sGDML PES and PIP PES. }
    \label{tab:freq_mode_ts2}
\end{table}

\newpage
\subsection{Timing Comparisons}
Having established that the sGDML and PIP PES provide precise fits to the CCSD(T) dataset, we consider the speed of evaluation of the PESs.  The timing was done on the same workstation and, in both cases, using Fortran 90 software.  Results for 100 000 evaluations of energy and energy plus gradient are given in Table \ref{tab:time}, relative to the PIP time for energy only.  First, note that sGDML is trained only for gradients, and since the energy is obtained from gradients, we leave the entry blank for the sGDML energy. The timing for the energy plus gradient is 13$\times$ the time for energy only for the PIP PES.  This is as expected for a standard analytical (forward) gradient evaluation as is done in MSA.  Note that this time is much faster (a factor of 15.5) than the time for the sGDML PES. 
As described in detail elsewhere\cite{PIPSJCP22,PESPIP}, fast reverse differentiation has been implemented in the Fortran software via a Mathematica script. As seen, there is a substantial speedup in the gradient evaluation (roughly a factor of 4).  Thus, the final ratio between the sGDML:PIP timing for energy plus gradient is 69.  For energy evaluation only relevant to quantum calculations, including the diffusion Monte Carlo (DMC) ones reported below, the factor is 206.  We discuss this large difference in speed (in line with a similar factor for ethanol\cite{PIPSJCP22}) below, where we summarize the assessments of the two fitting methods as applied here.  Before doing that, we present some results from a diffusion Monte Carlo calculation of the ground state wavefunction.
\begin{table}[]
\renewcommand{\arraystretch}{1.5}
    \centering
\begin{tabular}{l  r r} 
    \hline
     Time taken (s)    & sGDML  & PIP  \\
     \hline \hline
  Energy   &  - &  1  \\
  Energy + gradient &  206  & 13  \\
  Energy + fast reverse derivative & & 3 \\
  \hline
    \end{tabular}
\caption{Time per 100 000 calls}
    \label{tab:time}
\end{table}

\subsection{Diffusion Monte Carlo calculations of the zero point wavefunction}
Next, we present the results obtained through the DMC calculations~\cite{anderson1975random,kosztin1996introduction,mccoy2006diffusion} using our in-house software, as described in our recent paper on using DMC to locate ``holes" in a PES.\cite{dmcholes}  For each PES, we performed five DMC calculations initiated at the global minimum, with 20 000 random walkers and an imaginary time step of $\Delta \tau=5$ au for 30 000 time steps.  Upon completion of the unconstrained DMC, both PIP and sGDML PESs are identified as ``hole-free" surfaces, signifying the absence of any configuration with unphysical negative energies. However, the PIP PES exhibits a speed advantage in this computation, performing 360 times faster than the sGDML PES. The average ZPE obtained from five independent DMC simulations for PIP  and sGDML PES are $6641 \pm 2$ cm$^{-1}$ and $6612\pm 4$ cm$^{-1}$, respectively. These ZPEs are substantially lower than the harmonic ZPEs of 6965 and 6983 cm$^{-1}$, respectively, for the PIP and sGDML PESs.

\begin{figure}
    \centering
    \includegraphics[width=\textwidth]{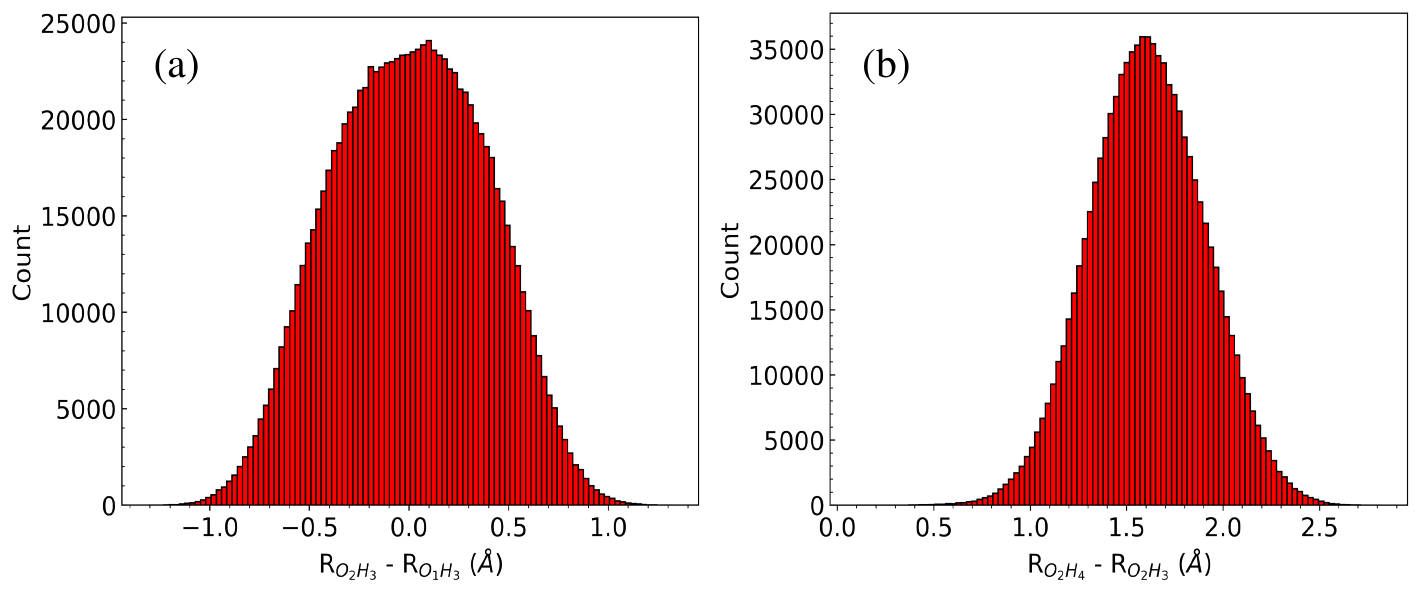}
    \caption{Histogram of cuts of the ground state DMC wavefunction vs the difference of the bond lengths of (a) R$_{O_2H_3}$ and R$_{O_1H_3}$ and (b) R$_{O_2H_4}$ and R$_{O_2H_3}$ (in \r{A}).  }
    \label{fig:dmc_bondlen_diff}
\end{figure}

We now present several plots of 1d wavefunctions from the DMC wavefunction obtained with the PIP PES.  These 1d wavefunctions are obtained from histograms of walkers for selected variables at the last time step of the DMC trajectory. Of major interest is the wavefunction of the shared H-atom, H$_3$ in Fig. \ref{fig:saddle_config}. This is shown using the difference variable $R_{\ce{O2H3}}$-$R_{\ce{O1H3}}$ in Figure~\ref{fig:dmc_bondlen_diff}, panel (a).  As seen, the peak is at zero and this signifies an equal sharing of the hydrogen atom between the two oxygen atoms, and consequently corresponds to the H-transfer TS. This is in agreement with earlier DMC studies of the ground state wavefunction using an earlier PIP PES\cite{mccoy2005quantum}.  As noted there this symmetric delocalization of the shared H-atom is due to the low potential energy (ca 80 cm$^{-1}$) of the H-transfer TS. This delocalization was noted in prior path integral \cite{Tuckerman1997, tuckerman2002nature, Tachikawa2005, Suzuki2008, arandhara24}, DMC \cite{mccoy2005quantum}, and MCTDH \cite{yang2008full} studies. 

\begin{figure}
    \centering
    \includegraphics[width = \textwidth]{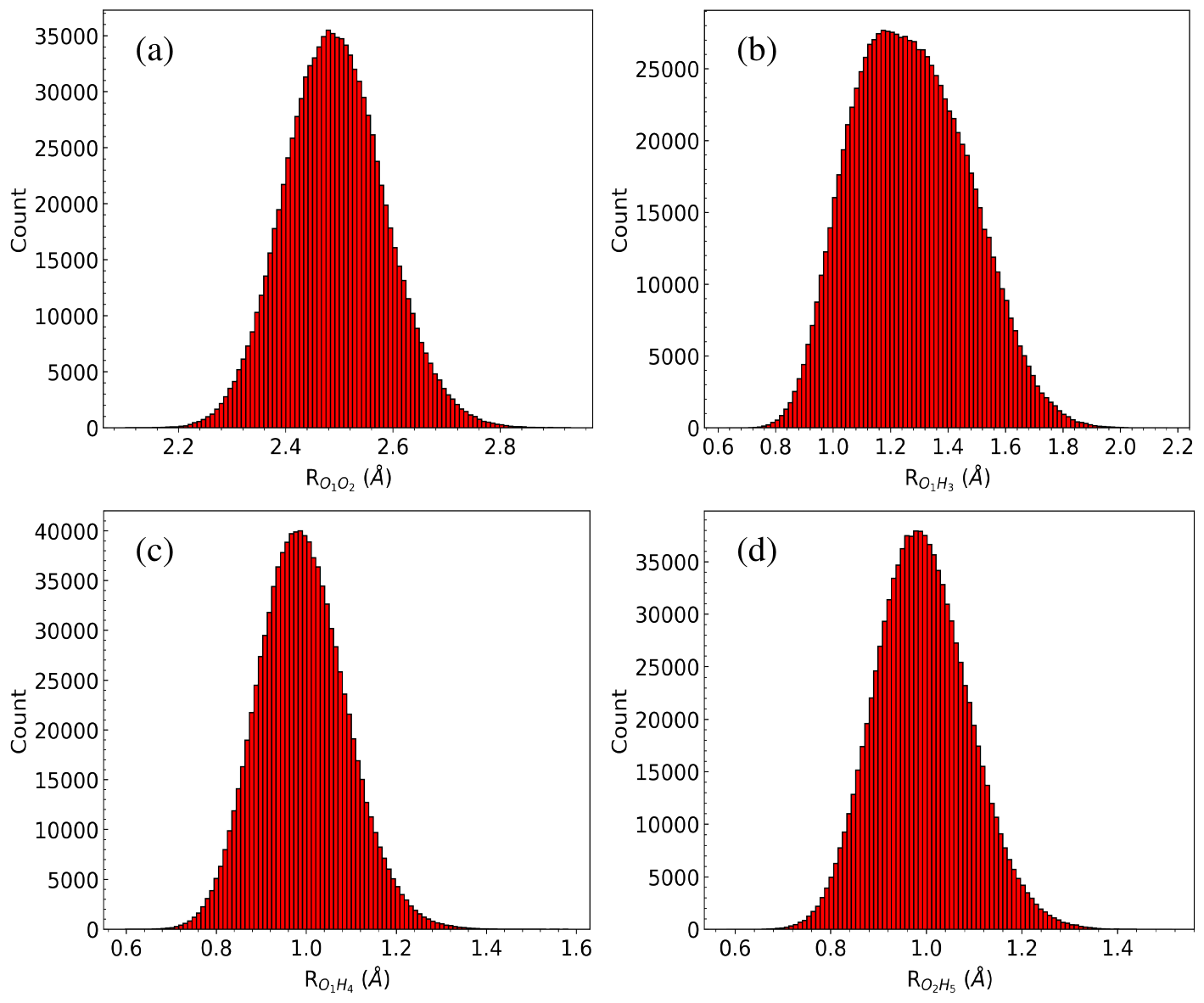}
    \caption{Histogram of cuts of the ground state DMC wavefunction vs the bond lengths of \ce{H3O2-} system (in \r{A}): (a) R$_{O_1O_2}$, (b) R$_{O_1H_3}$, (c) R$_{O_1H_4}$, and (d) R$_{O_2H_5}$, }
    \label{fig:dmc_bondlength}
\end{figure}

Next, we investigate a 1d wavefunction that provides information about the bifurcation TS. This is shown in Figure~\ref{fig:dmc_bondlen_diff}(b) in the difference variable $R_{\ce{O2H4}}-R_{\ce{O2H3}}$.  This variable is zero at the bifurcation TS.  As seen, unlike the result in panel (a), the histogram peaks at $1.58$ \r{AA}. This difference indicates that one hydrogen atom of the water molecule occupies the space between two oxygen atoms, while the other hydrogen atom remains farther away from the oxygen atom of \ce{OH-}. Note that the difference of the bond lengths  between \ce{O2H4} and \ce{O2H3} for the global minimum and the H-transfer TS are $1.42$\r{AA} and $1.59$ \r{AA}, respectively. On this plot the wavefunction is essentially zero at the bifurcation TS. This is not surprising given that the potential energy of this TS is roughly 2500 cm$^{-1}$. Finally, we show 1d wavefunctions for the indicated bond lengths in
Figure~\ref{fig:dmc_bondlength}.  Panel (a) shows the expected Gaussian shape for the O-O bond length extended over a range of 0.6 \r{AA}.  Panel (b) shows the large range of the shared H-atom motion with respect to the O atom. The wavefunctions in panels (c) and (d) for the \ce{O1H4} and \ce{O2H5}, respectively, are virtually identical, as expected from the structure in panel (c) of Fig. \ref{fig:saddle_config}. 


\newpage
\section{Summary and Conclusions}
We presented assessments of two potential energy surfaces of \ce{H3O2-}. One is a symmetric gradient domain machine learning (sGDML) PES, and the other is a new permutationally invariant polynomial (PIP) one.  These are successors to an earlier PIP potential energy surface (PES) reported in 2004.  We described the details of both fitting methods and then compared the two PESs with respect to precision, properties, and speed of evaluation. The two methods approach training differently. sGDML uses a subset of the data, which are exclusively 45 000 gradients. While this conforms to the usual train-test protocol, it is numerically necessary for sGDML and all kernel methods, which scale steeply in cost with the training data size.  PIPs use all the data, consisting of 15 024 energies and gradients, for a total data size of 240 384. This data size is easily managed in the least-squares fitting done in the PIPs approach. 

The two PESs are similarly precise; however, the PIP PES is much faster to evaluate for energies and energies plus gradient than the sGDML one, with factors of 200 and 70, respectively.  This factor of roughly two orders of magnitude is consistent with a similar factor found in a previous assessment for ethanol\cite{PIPSJCP22} and a very new one for 21-atom aspirin.\cite{houston2024headache}  Diffusion Monte Carlo calculations of the ground vibrational state wavefunction were done using both PESs.  Since these require just energies, the calculation using the PIP PES took roughly 300 times less CPU time than the sGDML one.  Analysis of the DMC wavefunction from the PIP PES calculation indicates that the shared proton is symmetrically located between OH groups but has near zero amplitude at the bifurcation saddle point. As noted previously\cite{huang2004quantum} and also recently\cite{arandhara24}, the bifurcation TS energy is much higher than the H-atom transfer one. Here, we see that this results in delocalization of the shared H-atom between two OH groups but significant localization of the H-atom with respect to the bifurcation pathway. This is consistent with results using PIMD simulations using the sGDML PES at low temperatures.\cite{arandhara24}  It would be interesting to investigate tunneling splittings, where the delocalized shared H-atom is replaced by a different H-atom via the bifurcation pathway.  The present fast PIP PES should enable such possible future studies.

\begin{acknowledgement}

JMB thanks the Army Research Office, DURIP grant (W911NF-14-1-0471), for funding a computer cluster where most of the calculations were performed. JMB and PP acknowledge current support from NASA grant (80NSSC22K1167). MA and SGR thank the Science and Engineering Research Board of India (grant EMR/2017/003881) for support with computational resources.
\end{acknowledgement}

\begin{suppinfo}
The Cartesian coordinates of the various stationary points of \ce{H3O2-}, optimized with the PIP PES, are provided.
\end{suppinfo}

\bibliography{ref}

\end{document}



\begin{table}[]
\renewcommand{\arraystretch}{1.5}
    \centering
    \begin{tabular}{l r r r r r r r r r }
        \hline
        \multirow{2}{*}{Atom} & \multicolumn{3}{c}{Global Minimum} & \multicolumn{3}{c}{TS Bifurcation}  & \multicolumn{3}{c}{TS H-transfer}\\
              \cmidrule(lr){2-4}  \cmidrule(lr){5-7}   \cmidrule(lr){8-10}
            & x & y & z & x & y & z & x & y & z\\
        \midrule
        O1  &  -1.2389 &  -0.0503 &  -0.0299 &  -1.3307 &  -0.0003 & -0.0014 & -1.2151 & -0.0516 & -0.0359\\
        O2  &   1.2425 &   0.0454 &  -0.0389 &   1.2807 & 0.0018   & -0.0352 & 1.2134  & 0.0501  & -0.0385\\
        H3  &  -0.1541 &  -0.0346 &  -0.0431 & -0.6476  & -0.6917  & 0.0243  & -0.0008 & -0.0007 & -0.0424\\
        H4  &  -1.4424 &   0.7066 &   0.5222 & -0.6487  &  0.6920  & 0.0274   & -1.4629 &  0.6590 & 0.5598 \\
        H5  &   1.5342 &  -0.6161 &   0.5944 & 1.9834   & 0.0009   & 0.6214  & 1.4626  & -0.6613 &  0.5557\\    
        \hline
    \end{tabular}
    \caption{Cartesian coordinates of the optimized global minimum, bifurcation TS, and H-transfer TS (in Angstrom), obtained from the PIP PES trained on all data.}
    \label{tab:SI_Coordinate1}
\end{table}

\begin{table}[]
\renewcommand{\arraystretch}{1.5}
    \centering
    \begin{tabular}{l r r r r r r  }
        \toprule
        \multirow{2}{*}{Atom} & \multicolumn{3}{c}{TS $cis$} & \multicolumn{3}{c}{TS $trans$} \\
              \cmidrule(lr){2-4}  \cmidrule(lr){5-7}  
          & x & y & z & x & y & z \\
        \midrule
         O1  &  -1.2471 &   -0.0515&   -0.0011& -1.2435  & -0.0641  &  0.0018\\
         O2  &   1.2528 &   -0.0542&    0.0017&  1.2434  &  0.0618  &  0.0013\\
         H3  &  -0.1722 &   -0.0427&    0.0001& -0.1637  &  -0.0456  &  0.0016\\
         H4  &  -1.4673 &    0.8813&   -0.0014& -1.4689  &  0.8676  &  0.0009\\
         H5  &   1.5582 &    0.8574&    0.0019&  1.6008  & -0.8299  &  0.0022\\
         
        \bottomrule
    \end{tabular}
    \caption{Cartesian coordinates of the optimized $cis$ and $trans$ \ce{HO-OH} torsion barriers (in Angstrom), obtained from the PIP PES trained on all data.} 
    \label{tab:SI_Coordinate2}
\end{table}